\documentclass[twocolumn,showpacs,amsmath,amssymb]{revtex4}

\usepackage{graphicx} 
\usepackage{amssymb}
\usepackage{amsfonts}
\usepackage{amsmath}

\begin{document}

\title{Predicting link directions via a recursive subgraph-based ranking}

\author{Fangjian Guo}
\author{Zimo Yang}
\email{yangzimo415@gmail.com}
\author{Tao Zhou}


\affiliation{Web Sciences Center, University of Electronic Science and Technology of China, Chengdu 611731, P. R. China}


\begin{abstract}
Link directions are essential to the functionality of networks and their prediction is helpful towards a better knowledge of directed networks from incomplete real-world data. We study the problem of predicting the directions of some links by using the existence and directions of the rest of links. We propose a solution by first ranking nodes in a specific order and then predicting each link as stemming from a lower-ranked node towards a higher-ranked one. The proposed ranking method works recursively by utilizing local indicators on multiple scales, each corresponding to a subgraph extracted from the original network. Experiments on real networks show that the directions of a substantial fraction of links can be correctly recovered by our method, which outperforms either purely local or global methods.
\end{abstract}

\maketitle

\section{Introduction}

Networks provide a powerful abstraction for describing the structures of a wide range of complex systems  \cite{Albert2002, Newman2003}. Among them, many belong to the class of directed networks --- a set of nodes connected by links, where each link is associated with a direction pointing from one node to another. The directions of links reflect the logical order of interaction or dependence between two nodes. For example, they  indicate the directional trend of information diffusion in an email network \cite{Eckmann2004} and the relations between leaders and followers in Twitter \cite{Kwak2010}. Other cases include the dependence of chemical substances in protein networks \cite{Wagner2001}, the preying relations among animals \cite{Pimm2002}, the hyperlinks connecting web pages \cite{Pastor-Satorras2007}, etc. Directions are essential to the functionality of networks: directness introduces asymmetric interactions into percolation and epidemic spreading on networks \cite{Bogun'a2005, Meyers2006}; directionality also influences the global emergence of collective behaviors \cite{Park2006} and is critical for synchronization in networks \cite{Son2009, Zhou2010, Zeng2011}.

Unfortunately, data collected from real networks are often incomplete, giving rise to the study of link prediction, which seeks to predict missing links according to the observed data \cite{Lue2011}. While in the simple case of undirected networks, only the possible existence of a link between two nodes $i$ and $j$ is concerned, the task is more complicated for directed networks, where the issue of existence and the issue of direction can be considered either simultaneously or separately: when simultaneously, one examines the existence of both $i \rightarrow j$ and $j \rightarrow i$; when separately, one first predicts whether a link, regardless of its direction, exists between $i$ and $j$ and then, if it exists, tries to determine the direction of that link ($i \rightarrow j$, $j \rightarrow i$ or bidirectional). Whereas previous works on predicting directed links generally follow the former scheme by fitting a statistical graph model \cite{Clauset2008}, using local motifs \cite{Zhang2012}, etc., we take the latter scheme in this paper. Specifically, while existence prediction can be aided by many similarity-based algorithms \cite{Liben-Nowell2007, Zhou2009}, we only focus on the essential problem of direction prediction, which remains largely to be investigated.

Assuming two nodes are connected, how to predict the direction between them? To answer this question, we seek to construct an optimal ordering of nodes such that a link tends to stem from a node with lower ranking and point to one with higher ranking. Admittedly, such a ranking-based method inevitably has its drawbacks mainly due to directed cycles, as real networks are usually not directed acyclic graphs (DAG). The desired property that a link points from a lower ranked node to a higher ranked one must be violated at least once for each directed cycle. And specifically, this suggests that the method cannot predict bidirectional links as they are simply directed cycles of length 2. Nevertheless, this method has its unique virtues: (i) it further reveals the potential functionality of ranking algorithms as a tool for investigating structural properties of networks, beyond their traditional role in information retrieval; (ii) we obtain both the predicted directions and a global ranking describing the directionality of the whole network, bridging the properties on both microscopic and macroscopic scales;  (iii) it may also serve as an effective approximation algorithm for \textit{linear ordering problem} and \textit{maximum acyclic subgraph problem} on directed networks, which are generally NP-hard and have been studied especially for tournament graphs (every pair of vertices is connected by a single directed link) \cite{Ailon2008}.

The rich structural information woven by directed links have motivated a number of ranking algorithms for information retrieval. They are designed to derive an ordering of nodes by leveraging the topological relations in the network and the ranking criteria is usually based on a global score. For example, PageRank \cite{Brin1998} ranks nodes by the stationary distribution of the probability of visitation by a random walker mimicking the behavior of an Internet surfer. Whereas PageRank powers the search engine of Google, its variants have also been applied to assessing the leadership in social networks \cite{Lue2011a}, the prestige of journals \cite{Bollen2006}, the ranking of scientists \cite{Radicchi2009} and their papers \cite{Chen2007}. Besides, HITS is another famous ranking algorithm that derives the ranking by a process of mutual recursion \cite{Kleinberg1999}. It defines two scores for each node, namely hub and authority. And a node with a high hub score points to many good authorities while one with a high authority score receives links from many good hubs.

However, for the task of predicting the direction between two given nodes, a ranking completely based on global quantities or processes can hardly capture the local directionality. Therefore, local indicators, such as in-degree and out-degree, should be utilized by our ranking algorithm. But the effectiveness of purely local indicators are weakened by their limited scope of information --- degrees are only related to directly connected nodes while indirect relations are lost. Thus, local indicators must be combined and rearranged carefully to form a meaningful global ranking. In this paper, as inspired by the hierarchical nature of disparate complex networks \cite{Clauset2008, Ravasz2003, Nagy2010, Mones2012}, we propose a method that uses local indicators recursively on multiple scales, each of which corresponds to a subgraph extracted from the whole network. Although local quantities may only give a rough global sketch, they can reliably capture local properties. Therefore, they should play a more decisive role as the scale diminishes due to their increasing fineness for describing relations in locality. Apart from its predictive purpose, our method may also lead to a deeper insight into the directional and hierarchical organization of many real networks.

\section{Problem description} 

Given a directed network $G(V, E)$, where $V$ denotes the set of nodes and $E$ the set of links, the directions of a portion of links are unknown (denoted by the set $E_c$), and we are then asked to predict the directions of these links based on the existence and directions of other known links (denoted by $E_n = E - E_c$), possibly also using the existence of the links in $E_c$. 

Among the varieties of possible solutions, we specifically consider resolving this problem by constructing a special ranking $R$. Denoting the place of node $i$ in the ranking as $R(i)$ (a small $R(i)$ means a top ranking), then for any link in $E_c$ connecting $i$ and $j$, the link is predicted to be $ i \rightarrow j $ if $R(i) > R(j)$ or $j \rightarrow i $ if $R(i) < R(j)$. As any two nodes are assigned to different places in the ranking, predicting two-way links is not considered here.

\begin{figure}
\includegraphics[scale=0.9]{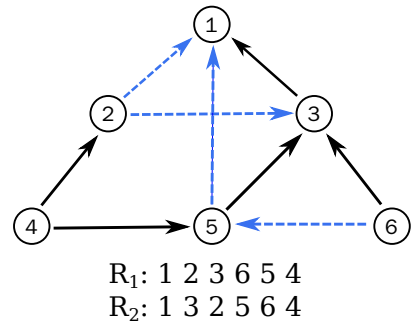}
\caption{A simple network where blue dashed links belong to $E_c$. Two rankings $R_1$ and $R_2$ are given below the network. Ranking $R_1$ reaches a conformity of $0.5$ by giving two opposite predictions (contradicting $2 \rightarrow 3$ and $6 \rightarrow 5$), whereas $R_2$ reaches a perfect conformity of 1. }
\label{fig:schematic_compatibility}
\end{figure}

Once the directions of the links in $E_c$ are discovered, the performance of a ranking $R$ can be evaluated by computing its conformity with these links, i.e. the accuracy of direction prediction, given by
\begin{equation}
	C = \frac{\Vert \{ (i, j) \in E_c \mid R(i) > R(j), i \rightarrow j \} \Vert}{\Vert E_c \Vert },
\end{equation}
where $\Vert \cdot \Vert$ denotes the number of elements in the set and $(i,j)$ denotes a link between $i$ and $j$ (both $i \rightarrow j$ and $j \rightarrow i$ are counted if $i$ and $j$ are found to be reciprocally connected). $C$ is simply the ratio of correctly predicted directions to the total number of links in $E_c$. The maximum value of $C$ is 1 corresponding to a perfect prediction, although not always attainable due to cycles in the network, and a value of 0.5 means guessing the direction by pure chance. Fig. \ref{fig:schematic_compatibility} gives a simple example where $R_1$ reaches a conformity 
of 0.5 and $R_2$ reaches a perfect conformity of 1.

\section{Methods}

Our method relies on the assumption that the formation of networks is regulated by an implicit ranking of nodes, such that links tend to originate from lower-ranked nodes and point to higher-ranked ones. Such a ranking, if can be approximately derived from the observed data, is therefore useful for predicting the directions of missing links. While a maximum-likelihood method has been recently proposed for extracting this ranking from friendship networks \cite{Ball2012}, we take a different approach by combining local indicators with hierarchical organizations of networks.

\begin{figure}
\includegraphics{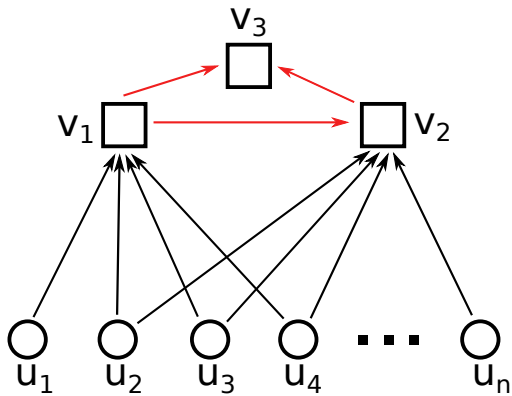}
\caption{A network made up of followers $u_1, u_2, u_3, \cdots, u_n$ and a few leaders $v_1, v_2, v_3$. While the stratification between leaders and followers can be determined by degree differences, the ordering among leaders has to be obtained by extracting their induced subgraph (red links).}
\label{fig:schematic_scale}
\end{figure}

Our method is best explained by considering a simple example of social networks as illustrated by Fig. \ref{fig:schematic_scale}, which is made up of a few leaders $v_1, v_2, v_3$ and many followers $u_1, u_2, \cdots, u_n$. Intuitively, leaders should enjoy higher ranking than followers and local quantities like degrees are useful for identifying both of them. As leaders are supposed to have bigger in-degrees and smaller out-degrees, we adopt the the degree difference $D^\Delta = D^{\text{in}} - D^{\text{out}}$ as the local indicator, which, as will be demonstrated by experiments, outperforms either in-degree or out-degree alone. Clearly from the example, the leaders $v_1, v_2, v_3$ can be separated from followers by noting their higher degree differences ($D^\Delta$ is positive for $v_1, v_2, v_3$ while negative for $u_1, u_2, \cdots, u_n$).

However, the internal relations among leaders cannot be readily determined in this way as the large number of their followers may overwhelm the degrees induced by their interrelations. In this case, although $v_2$ has the highest degree difference ($D^\Delta = 4$), while $v_1$ and $v_3$ have the same lower degree difference ($D^\Delta = 2$), $v_3$ is obviously the leader of the highest level. This problem can be remedied by leveraging the hierarchical nature of disparate networks on multiple scales \cite{Ravasz2003, Clauset2008, Nagy2010, Mones2012} --- the relations among nodes on a smaller scale can be determined in a way similar to that on a larger scale. We then explore the relations among $v_1, v_2$ and $v_3$ by extracting the subgraph induced by them and their degree differences in the subgraph evidently reveal their ordering --- $v_3, v_2, v_1$. All of them, of course, are placed higher than $u_1, u_2, \cdots, u_n$ in the global ranking.

We develop this idea into our algorithm: on a certain scale, nodes in the graph are divided into two classes by sorting the degree difference of each node. The internal orderings of nodes in each class are respectively determined in the subgraphs induced by them in a recursive fashion, while always placing the class of leaders as a whole ahead of the class of followers in the ranking.

A detailed explanation is presented as follows. Considering a directed network $ G(V,E) $, we examine the network on a certain scale by focusing on the subgraph $G_{\widetilde V}$ induced by a subset of nodes $\widetilde V \subseteq V $. For any subgraph $G_{\widetilde V}$ (including G itself), let $I(i; \widetilde V) = l$ ($ 1 \leq l \leq \Vert \widetilde V \Vert $) denote that node $i$ ($ i \in \widetilde V $) takes the $l$-th place sorted by the degree difference $D^\Delta$ in the subgraph in descending order. Then the set of nodes $ \widetilde V $, if large enough, is further divided into the set of leaders $ V_L(\widetilde V) $ and the set of followers $V_F (\widetilde V)$ based on this order, while assuming a factor $ \alpha $ ($ 0 < \alpha < 1 $) controlling the relative size of each, given by
\begin{equation}
	V_L(\widetilde V) = \{ j \in \widetilde V | I(j;\widetilde V) \leq \alpha \Vert \widetilde{V} \Vert \},
\end{equation}
\begin{equation}
	V_F(\widetilde V) = \{ j \in \widetilde V | I(j;\widetilde V) > \alpha \Vert \widetilde{V} \Vert \}.
\end{equation}

\begin{figure}
\centerline{\includegraphics[scale=0.65]{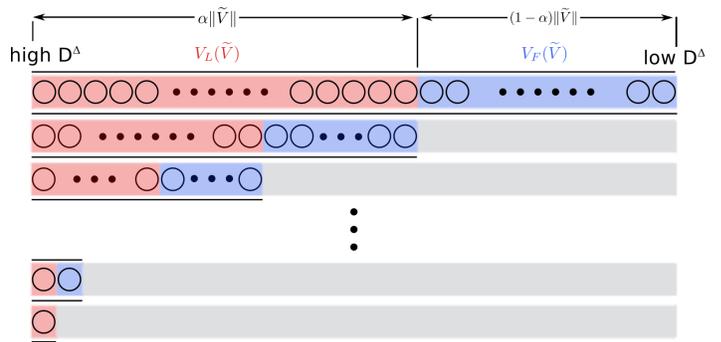}}
\caption{A schematic illustration of the recursive ranking procedure, where the red section denotes the set of leaders and the blue section denotes the set of followers. Such division after reordering occurs on consecutively diminishing scales, until the subgraph contains too few nodes for subdivision. Here we focus on the process within the initial class of leaders and other nodes are masked by grey shades for clarity.}
\label{fig:schematic_ranking}
\end{figure}

The relative ranking of node $i$ with respect to $G_{\widetilde V}$ is defined recursively as
\begin{eqnarray}
R(i;\widetilde V) = 
	\begin{cases}
		I(i; \widetilde V) & \Vert \widetilde{V} \Vert < \frac{1}{\alpha}\\
		R(i; V_L(\widetilde V)) &  \Vert \widetilde{V} \Vert \geq \frac{1}{\alpha}, i \in V_L(\widetilde V) \\
		\Vert V_L(\widetilde V) \Vert + R(i; V_F(\widetilde V)) & \Vert \widetilde{V} \Vert \geq \frac{1}{\alpha}, i \not\in V_L(\widetilde V),
	\end{cases}
\end{eqnarray}
where the first case corresponds to the triviality that $\Vert \widetilde V \Vert$ being too small for subdivision, while the second and third correspond to the node being a leader and being a follower on a smaller scale respectively. If it is a leader, its place compared with other leaders is simply used; if a follower, we also need to add the total number of leaders in $V_L$ to its place among other followers, due to the rule that followers are always ranked behind leaders as a whole. Such recursive reordering and division occurring on consecutively diminishing scales is schematically illustrated by Fig. \ref{fig:schematic_ranking}. For example, for a network with $N = 10,000$ nodes and $\alpha = 0.6$, the set of all nodes is firstly divided into $6,000$ leaders and $4,000$ followers, and then the $6,000$ leaders are further divided into $3,600$ leaders and $2,400$ followers on the next scale. Such recursive division continues until only one node is left in the subgraph.

Finally, the ranking derived by our method is simply given by 
\begin{equation}
	R(i) = R(i;V),
\end{equation}
where $V$ is the collection of all nodes in the entire network.

All links in $E_c$ are removed before applying this ranking method to the network (preserving these links as virtual two-way links will produce the same result due to cancellation in $D^\Delta$). Direction prediction is simple once the ranking is obtained: each link in $E_c$ is predicted to be pointing from the lower-ranked node to the higher-ranked node.

\section{Experiments}
By using data of real networks, our ranking method is parameterized by selecting an optimal $\alpha$. Then its performance for predicting direction of links is compared to ranking by in-degree, out-degree, degree difference $D^\Delta$ as well as PageRank.

Four real networks are used for experiments: (i) Gnutella P2P network \cite{Leskovec2007, Matei2002}, the peer-to-peer file sharing network of Gnutella, where one host is connected to another by a directed link; (ii) Facebook wall posts network \cite{Viswanath2009}, the network formed by wall posts of Facebook users in New Orleans, where a link from user $A$ to user $B$ means $A$ has posted on $B$'s wall; (iii) Slashdot zoo network \cite{Kunegis2009}, the social network of slashdot.org, where a link from user $A$ to user $B$ means $A$ has endorsed $B$ as either ``friend'' or ``foe''; (iv) C. elegans neural network \cite{White1986, Watts1998}, the neural network of the worm C. elegans, where a directed link corresponds to a chemical synapse along which signals can be passed from one neuron to another. Their sizes are presented in Table \ref{tab:data_description}. Note that the largest weakly connected component is used here for Gnutella P2P network and the Facebook wall posts network as they are not connected. Multiple links and self-loops are removed if contained in the original network.

\begin{table}
\caption{Dataset description}
\label{tab:data_description}
\begin{center}
\begin{tabular}{lll}
\hline
Dataset & $\Vert V \Vert$ & $\Vert E \Vert$ \\
\hline
Gnutella P2P network & 8,104 & 26,008 \\
Facebook wall posts network & 43,953 & 262,631 \\
Slashdot zoo network & 79,120 & 515,571 \\
C. elegans neural network & 297 & 2,345 \\
\hline
\end{tabular}
\end{center}
\end{table}

As $\alpha$ specifies the relative size of $V_L(\widetilde V)$ and $V_F(\widetilde V)$, we seek to select its value by examining the ranking's global conformity with all one-way links (with no reverse link) by applying our method to the whole network. Denoting the set of all one-way links by $E_g = \{ \langle i, j \rangle \in E \mid \langle j, i \rangle \not\in E \}$, where $\langle i, j \rangle$ refers to a link from $i$ to $j$, then the global conformity is given by
\begin{equation}
	C_g = \frac{\Vert \{ \langle i, j \rangle \in E_g \mid R(j) < R(i) \} \Vert}{\Vert E_g \Vert},
\end{equation}
which is the ratio of one-way links whose directions are in agreement with the ranking to the total number of one-way links.

\begin{figure}
\includegraphics[scale=0.5]{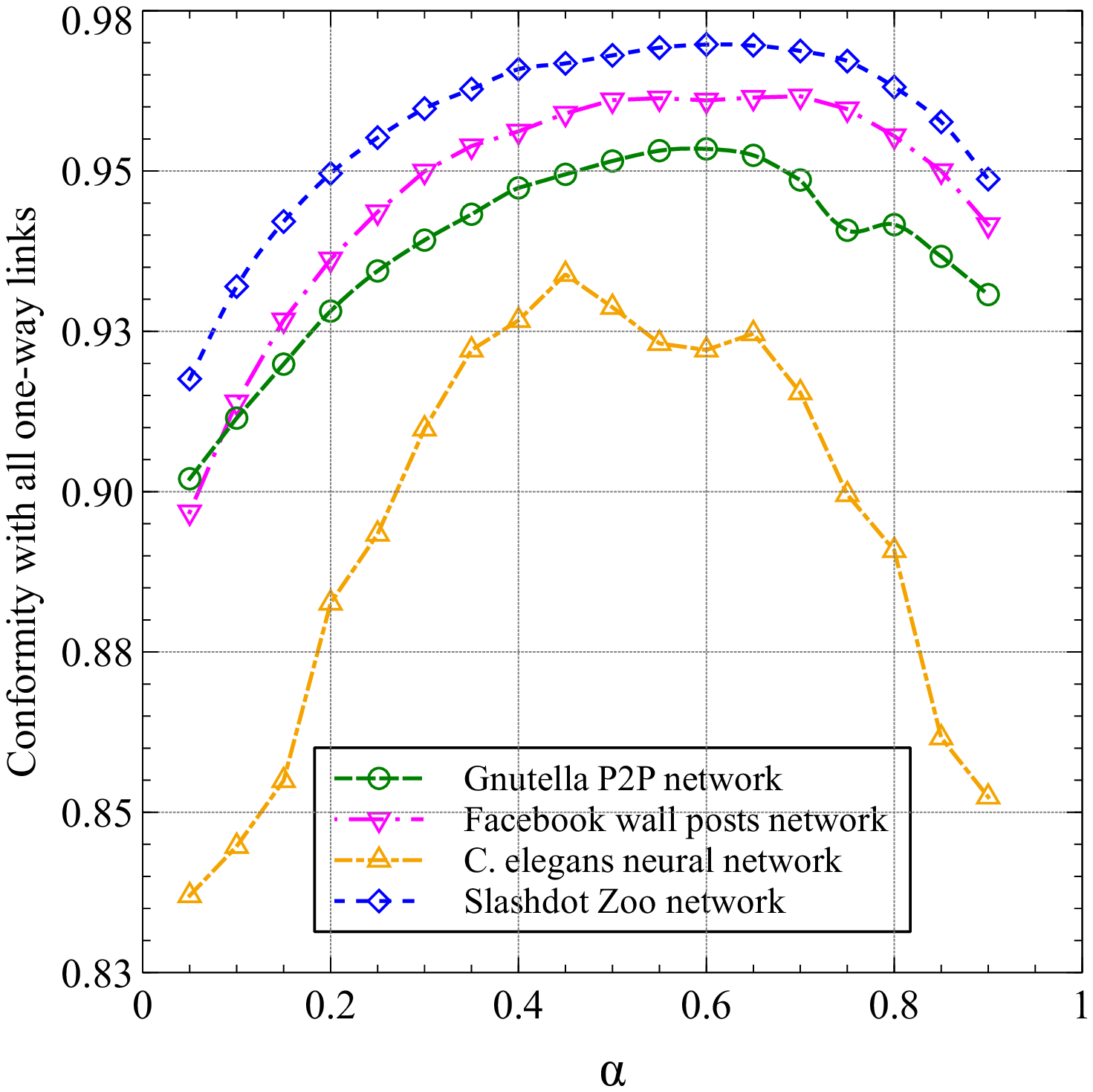}
\caption{The global conformity $C_g$ with one-way links versus different values of $\alpha$. The optimal value for $\alpha$ occurs around 0.6 for most networks.}
\label{fig:alpha_change}
\end{figure}

Figure \ref{fig:alpha_change} reports the global conformity $C_g$ under different values of $\alpha$. It is found that the optimal value for $\alpha$ that maximizes conformity lies around $0.6$ for tested networks except the neural network of C. elegans. $C_g > 0.92$ is reached at $\alpha = 0.6$ for all networks, indicating the ranking's high conformity with link directionality. Therefore, for simplicity, we choose $\alpha = 0.6$ for the task of direction prediction.

We randomly select a portion of links (denoted by the set $E_c$) out of all one-way links $E_g$ in a real network and a ranking's performance for direction prediction is evaluated by computing the ranking's conformity with these links. Ranking algorithms are performed on the network after removing the links in $E_c$. Only one-way links are used for evaluation because any ranking would be half-right and half-wrong for a pair of nodes connected by reciprocal links by our criteria. Besides our method, four other ranking methods, with the same prediction rule from rankings, are used for comparison: (i) PageRank, (ii) ranking in descending order of in-degree, (iii) ranking in ascending order of out-degree and (iv) ranking in descending order of degree difference $D^\Delta$, which is the local indicator used in our method.

The algorithm of PageRank is briefly described as follows. The PageRank score of a node $i$ in the network can be computed by \cite{Page1999}
\begin{equation}
P_t(i) =  c\sum_{j:j \rightarrow i} \frac{P_{t-1}(j)}{D^{\text{out}}_j} + \frac{1-c}{N},
\end{equation}
where $P_t(i)$ denotes the probability of visiting node $i$ at the time step $t$ by a random walker. This random walker moves along the links of the network with probability $c$, corresponding to the first term in the right-hand side, while jumping to a randomly chosen node with probability $(1-c)$, corresponding to the second term. The damping factor $c$ is set to be $0.85$ as commonly used \cite{Brin1998} and we have tested that its performance as a direction predictor is insensitive to this factor. By computing the formula above iteratively, a steady state can be reached and all nodes are then ranked in descending order of the probability $P(i)$ in the stationary distribution.

\begin{figure}
\includegraphics[scale=0.185]{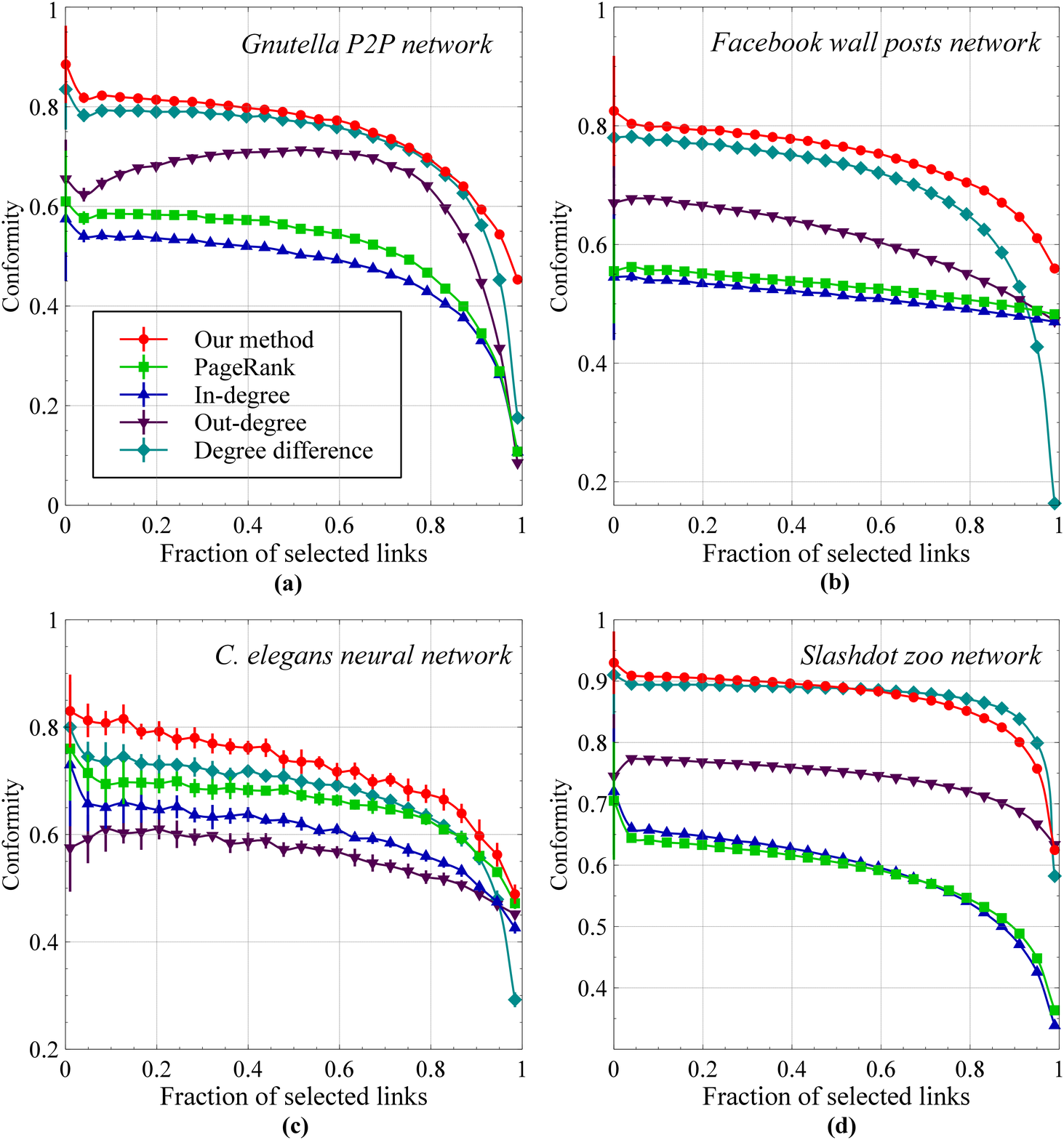}
\caption{The performance of our method compared with PageRank, in-degree, out-degree and degree difference $D^\Delta$ on real networks: (a) Gnutella P2P, (b) Facebook wall posts, (c) Slashdot zoo, (d) C. elegans. Conformity $C$ is drawn against the fraction of selected links among one-way links $\Vert E_c \Vert / \Vert E_g \Vert$. The results are obtained by averaging over 10 independent runs and error bars represent standard deviations, which may be too small to be seen in (a), (b) and (d).}
\label{fig:prediction_result}
\end{figure}

Figure \ref{fig:prediction_result} reports the results on four real networks, where conformity $C$ is drawn against the fraction of selected links among all one-way links $\Vert E_c \Vert / \Vert E_g \Vert$. Our method obviously outperforms other methods, achieving especially high conformity for networks of Gnutella P2P and Slashdot zoo, validating its effectiveness for direction prediction. The performance of our method is also stable and only small decrease in conformity is observed even when $\Vert E_c \Vert / \Vert E_g \Vert$ reaches 0.5.  Meanwhile, $D^\Delta$ is evidently a better local indicator for direction than $D^{\text{in}}$ and $D^{\text{out}}$. In fact, despite its simplicity, its performance approaches and even exceeds the performance of our method when many links are removed in Slashdot zoo network.

\section{Conclusion and discussion}
In directed networks, directions of links and rankings are closely connected. While directions provide rich topological information for ranking algorithms, a proper ranking of nodes also reflects the directional relations among nodes. In this paper we explore the latter aspect of this connection and we use the presented ranking method to predict unknown directions of links in a network, complementing current progress on the topic of link prediction. 

Directions are related to both local measures and global properties, where the trade-off between the two is a tough challenge. Purely relying on either local or global measures can hardly produce effective inference of the directions of links. This difficulty can be much resolved by considering the hierarchical structure of real networks. Simple local measures like in-degree and out-degree, despite their limited fineness at a global scale, tend to tell us more about the topology as we investigate the network at a smaller scale by extracting the subgraph induced by a fewer number of nodes. This procedure naturally goes in a recursive fashion as the hierarchical structure is in itself self-similar \cite{Ravasz2003, Song2005, Andrade2005, Zhou2005}. 

Apart from the purpose of direction prediction, our method can also be used as an effective heuristic for constructing the maximum acyclic subgraph of a directed network.

\acknowledgments
This work is partially supported by the National Science Foundation of China under Grant No. 11075031 and the Fundamental Research Funds for the Central Universities.

\bibliography{DriectionRanking}

\end{document}